\documentclass[reprint,twocolumn,superscriptaddress]{revtex4}%
\usepackage{graphicx}
\usepackage{bm}
\usepackage{latexsym}
\usepackage{epsf}
\usepackage{rotating}
\usepackage{epsfig,graphics,rotate,color}
\usepackage{wrapfig}
\usepackage{amssymb}
\usepackage{amsmath}
\usepackage{amsfonts}
\usepackage{subfigure}
\usepackage{array,hhline,dcolumn}%
\usepackage[normalem]{ulem}
\providecommand{\U}[1]{\protect\rule{.1in}{.1in}}
\begin{document}


\title{First Measurements of Inclusive Muon Neutrino Charged Current Differential Cross Sections on Argon}

\author{C. Anderson}
\affiliation{Yale University, New Haven, CT 06520}

\author{M. Antonello}
\affiliation{INFN - Laboratori Nazionali del Gran Sasso, Assergi, Italy}

\author{B. Baller}
\affiliation{Fermi National Accelerator Laboratory, Batavia, IL 60510}

\author{T. Bolton}
\affiliation{Kansas State University, Manhattan, KS 66506}

\author{C. Bromberg}
\affiliation{Michigan State University, East Lansing, MI 48824}

\author{F. Cavanna}
\affiliation{Universita dell'Aquila e INFN, L'Aquila, Italy}

\author{E. Church}
\affiliation{Yale University, New Haven, CT 06520}

\author{D. Edmunds}
\affiliation{Michigan State University, East Lansing, MI 48824}

\author{A. Ereditato}
\affiliation{University of Bern, Bern, Switzerland}

\author{S. Farooq}
\affiliation{Kansas State University, Manhattan, KS 66506}

\author{B. Fleming}
\affiliation{Yale University, New Haven, CT 06520}

\author{H. Greenlee}
\affiliation{Fermi National Accelerator Laboratory, Batavia, IL 60510}

\author{R. Guenette}
\affiliation{Yale University, New Haven, CT 06520}

\author{S. Haug}
\affiliation{University of Bern, Bern, Switzerland}

\author{G. Horton-Smith}
\affiliation{Kansas State University, Manhattan, KS 66506}

\author{C. James}
\affiliation{Fermi National Accelerator Laboratory, Batavia, IL 60510}

\author{E. Klein}
\affiliation{Yale University, New Haven, CT 06520}

\author{K. Lang}
\affiliation{The University of Texas at Austin, Austin, TX 78712}

\author{P. Laurens}
\affiliation{Michigan State University, East Lansing, MI 48824}

\author{S. Linden}
\affiliation{Yale University, New Haven, CT 06520}

\author{D. McKee}
\affiliation{Kansas State University, Manhattan, KS 66506}

\author{R. Mehdiyev}
\affiliation{The University of Texas at Austin, Austin, TX 78712}

\author{B. Page}
\affiliation{Michigan State University, East Lansing, MI 48824}

\author{O. Palamara}
\affiliation{INFN - Laboratori Nazionali del Gran Sasso, Assergi, Italy}

\author{K. Partyka}
\affiliation{Yale University, New Haven, CT 06520}

\author{A. Patch}
\affiliation{Yale University, New Haven, CT 06520}

\author{G. Rameika}
\affiliation{Fermi National Accelerator Laboratory, Batavia, IL 60510}

\author{B. Rebel}
\affiliation{Fermi National Accelerator Laboratory, Batavia, IL 60510}

\author{B. Rossi}
\affiliation{University of Bern, Bern, Switzerland}

\author{M. Soderberg}
\affiliation{Fermi National Accelerator Laboratory, Batavia, IL 60510}
\affiliation{Syracuse University, Syracuse, NY 13244}

\author{J. Spitz}
\affiliation{Yale University, New Haven, CT 06520}

\author{A.M. Szelc}
\affiliation{Yale University, New Haven, CT 06520}

\author{M. Weber}
\affiliation{University of Bern, Bern, Switzerland}

\author{T. Yang}
\affiliation{Fermi National Accelerator Laboratory, Batavia, IL 60510}

\author{G. Zeller}
\affiliation{Fermi National Accelerator Laboratory, Batavia, IL 60510}

\collaboration{The ArgoNeuT Collaboration}
\noaffiliation


\begin{abstract} 
The ArgoNeuT collaboration presents the first measurements of inclusive muon neutrino charged current differential cross sections on argon. Obtained in the NuMI neutrino beamline at Fermilab, the flux-integrated results are reported in terms of outgoing muon angle and momentum.  The data are consistent with the Monte Carlo expectation across the full range of kinematics sampled, $0^\circ$$<\theta_\mu$$<36^\circ$ and $0$$<P_\mu$$<25$~GeV/c. Along with confirming the viability of liquid argon time projection chamber technology for neutrino detection, the measurements allow tests of low energy neutrino scattering models important for interpreting results from long baseline neutrino oscillation experiments designed to investigate CP violation and the orientation of the neutrino mass hierarchy. 
\end{abstract}

\maketitle

Precision neutrino cross section measurements are required in order to fully characterize the properties of the neutrino-nucleus interaction and are important for the reduction of systematic uncertainties in long baseline neutrino oscillation experiments sensitive to non-zero $\theta_{13}$, CP-violation in the lepton sector, and the orientation of the neutrino mass hierarchy. A collection of inclusive muon neutrino charged current ($\nu_\mu$ CC) interactions can be considered a ``standard candle" for characterizing the composition of a neutrino beam as event identification is insensitive to the complicating effects of intra-nuclear effects and experiment-specific exclusive channel definitions. As such, CC-inclusive samples remain free from significant background contamination, regardless of the experimental configuration. Despite the preponderance of total cross section results, most recently in Refs.~\cite{flux,nomadcc,sciboonecc}, differential cross section measurements as a function of outgoing particle properties are sparse. Such measurements are necessary for obtaining a complete kinematic description of neutrino-nucleus scattering. This letter presents $\nu_\mu$ CC differential cross sections as measured with ArgoNeuT (Argon Neutrino Test) in a neutrino/muon kinematic range relevant for MINOS~\cite{minossteel}, T2K~\cite{t2knim}, NOvA~\cite{nova}, and LBNE~\cite{lbne}. The total $\nu_\mu$ CC cross section at $\langle E_\nu \rangle=4.3$~GeV is also reported.

ArgoNeuT is the first liquid argon time projection chamber (LArTPC)~\cite{rubbia} to take data in a low energy neutrino beam, and the second at any energy~\cite{wanfprd}. ArgoNeuT collected neutrino and anti-neutrino events in Fermilab's NuMI beamline~\cite{numibeam} at the MINOS near detector (henceforth referred to as ``MINOS") hall from September 2009 to February 2010. Along with performing timely and relevant physics, the ArgoNeuT experiment represents an important development step towards the realization of a kiloton-scale precision LArTPC-based detector to be used for understanding accelerator- and atmospheric-based neutrino oscillations, proton decay, and supernova burst/diffuse neutrinos. 

\begin{figure}[h!]
\begin{center}
\begin{tabular}{c}
\hspace{-.7cm}
\includegraphics[scale=.26]{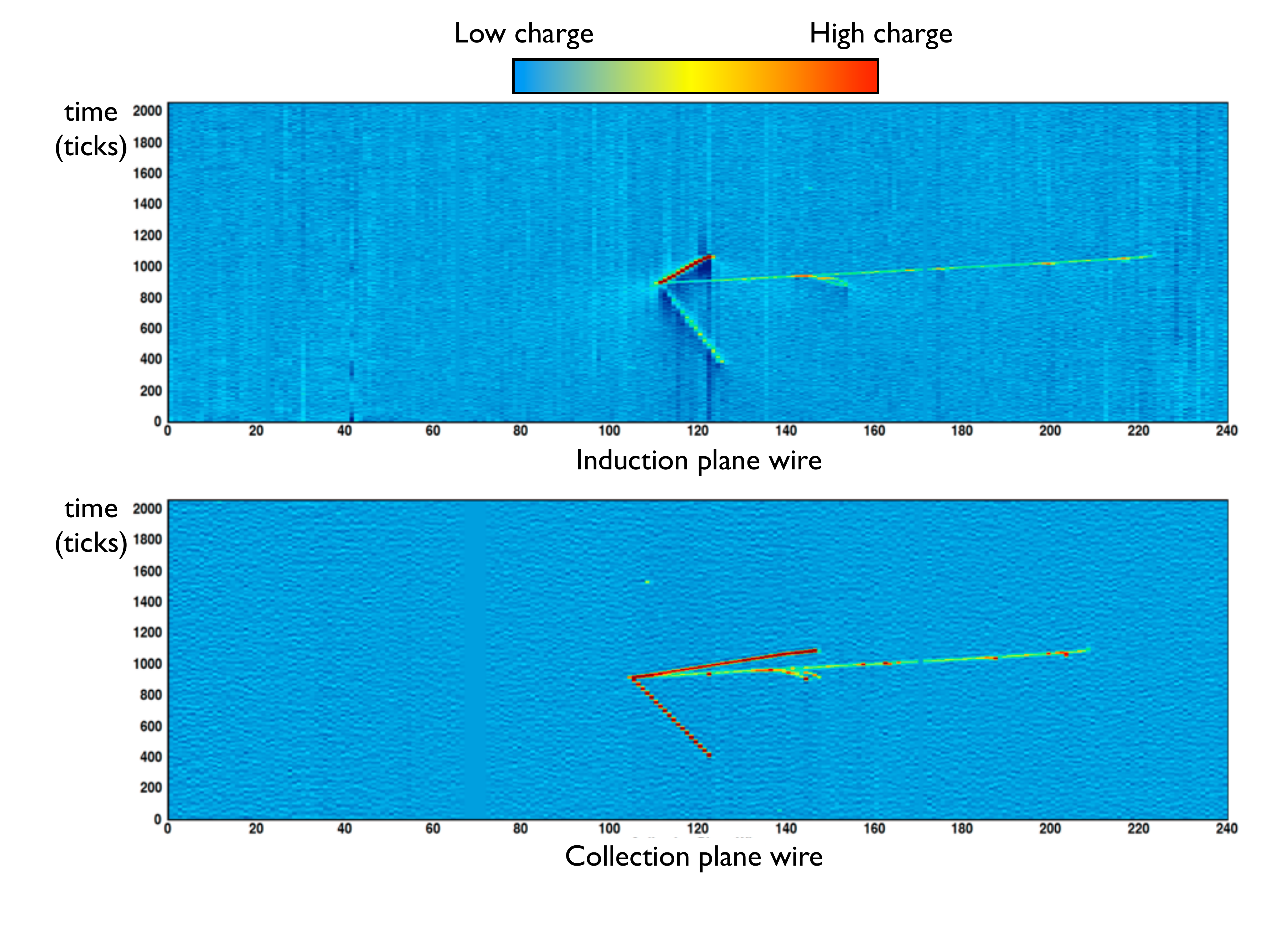}
\end{tabular}
\end{center}
\vspace{-1.cm}
\caption{A $\nu_\mu$ CC candidate event as seen in ArgoNeuT's induction and collection wire plane views. The color is representative of the amount of charge detected by the wires. There are 240~wires on each plane, the spacing between adjacent wires is 4~mm, and each time ``tick" corresponds to 198~ns.}
\label{fig:event}
\end{figure}

ArgoNeuT employs a set of two wire planes at the edge of a 170~liter TPC in order to detect neutrino-induced particle tracks. A 500~V/cm electric field imposed in the liquid argon volume of the TPC allows the ionization trails created by charged particles to be drifted toward the sensing wire planes. The ionization induces a current on the inner ``induction" wire plane as it approaches and recedes and is subsequently collected on the outer ``collection" wire plane. The signal information from the wire planes, oriented with respect to one another at an angle of 60$^\circ$, combined with timing provide a three dimensional picture of the neutrino event with complete calorimetric information~\cite{bromberg}.  Figure~\ref{fig:event} depicts a $\nu_\mu$ CC candidate event collected in the 47 $\times$ 40 $\times$ 90~cm$^3$ (drift $\times$ vertical $\times$ beam coordinate) ArgoNeuT TPC. 

The differential cross section in terms of a measured variable $u$ in bin $i$ is given by
\begin{equation}
\frac{\partial \sigma (u_{i})}{\partial u }=\frac{N_{\mathrm{measured},i}-N_{\mathrm{background},i}}{\Delta u_{i}~\epsilon_i~N_{\mathrm{targ}}~\Phi}~,
\label{diffeq}
\end{equation}
where $N_{\mathrm{measured},i}$ represents the number of signal and background events passing analysis selection, $N_{\mathrm{background},i}$ is the number of expected background events, $\Delta u_{i}$ is the bin width, $\epsilon_i$ is the detection efficiency, $N_{\mathrm{targ}}$ is the number of argon nucleus targets in the fiducial volume, and $\Phi$ is the total neutrino flux exposure. The variable $u$ is measured as outgoing muon angle with respect to the initial neutrino direction ($\theta_\mu$) and momentum ($P_\mu$) in this letter.

Neutrino event characterization takes place with the LArSoft~\cite{larsoft} automated reconstruction software. The software identifies hits, clusters proximal hits together, and identifies and fits line-like objects on each of the two wire plane views using a technique based on the Hough transform~\cite{hough}. Three dimensional track reconstruction proceeds with the union of line-like objects from each plane view that feature endpoints common in time. The process iterates until all line-like objects associated with an identified neutrino interaction vertex in the liquid argon volume are considered.  

As ArgoNeuT is too small to completely contain GeV-scale muons, muon momentum and charge are determined by MINOS. The front face of MINOS is approximately 1.5~m downstream of ArgoNeuT, and the center of ArgoNeuT is located 20~cm below the center of the MINOS fiducial volume. After track formation, an attempt is made to match the three dimensional tracks that leave the ArgoNeuT TPC with muons that have been reconstructed in MINOS and have a hit within 20~cm of the upstream face of the detector. The matching criteria are based on the radial and angular differences between the projected-to-MINOS ArgoNeuT track and the candidate MINOS track. In the case that a muon stops in MINOS, the track's complete energy deposition is used for the measurement. In the case that a muon is not contained, track curvature in the toroidal magnetic field of the detector is employed~\cite{minossteel}. The ArgoNeuT detector simulation is used to account for the muon energy lost before reaching MINOS, given the measured path length from the in-ArgoNeuT interaction vertex to the most upstream hit of the matched MINOS track ($\langle E_{\mathrm{lost}}\rangle$=200~MeV).

The detection efficiency, measurement resolution, and $\overline{\nu}_\mu$/neutral-current background estimates are obtained using the reconstruction software applied to simulated neutrino events. The simulation employs a GEANT4-based~\cite{geant4} detector model and particle propagation software in combination with the GENIE neutrino event generator~\cite{genie}. It incorporates the complete detector geometry including the ArgoNeuT TPC, cryostat, and containment vessel, along with induction/collection plane signal formation, electronic noise, electron lifetime, ionization diffusion, and electron-ion recombination. A full simulation of MINOS, as provided by the collaboration, is utilized as well.

A set of simple selection criteria is used to remove background events that mimic $\nu_\mu$ CC signal events in this analysis. Muons created in upstream neutrino interactions can enter ArgoNeuT and be reconstructed. Extended neutral current and mis-reconstructed $\overline{\nu}_\mu$ events can enter the sample as well. The neutrino event's vertex is required to be inside of the ArgoNeuT fiducial volume, 3~cm from the sense wires and cathode plane, 4~cm from the top and bottom of the TPC, 6~cm from the upstream end of the TPC, and 4~cm from the downstream end. The fiducial volume is defined this way in order to ensure that muons created in interactions upstream do not enter the signal sample and to allow a substantial track length in the active volume for effective reconstruction. The track matching criteria along with a requirement that the reconstructed and matched MINOS track is negatively charged represent the only other selection criteria used in this analysis. 

Corresponding to 8.5$\times$10$^{18}$~protons on target (POT) collected in low energy NuMI neutrino-mode, there are 373 and 362 $\nu_\mu$ CC-like events that enter the $0^\circ$$<\theta_\mu$$<36^\circ$ and $0$$<P_\mu$$<25$~GeV/c measurement ranges, respectively. The measurement ranges are chosen in consideration of sparsely populated bins and avoiding regions of low acceptance. The flux is reported in Table~\ref{fluxtable}. For the 3-50~GeV NuMI neutrino energy range, the flux prediction comes directly from Ref.~\cite{flux}. For the 0-3~GeV range, the flux prediction is determined using a Monte Carlo simulation of the NuMI beamline and is independent of MINOS neutrino data and cross section assumptions. The MINOS measured flux (3-50~GeV) and flux prediction (0-3~GeV) are consistent near the transition.  

After subtracting the expected 18 event background contribution, the selected $\theta_\mu$ and $P_\mu$ distributions are efficiency corrected on a bin-by-bin basis according to Eq.~\ref{diffeq}. A $\nu_\mu$ CC event that originates in the ArgoNeuT fiducial volume enters the signal sample after ArgoNeuT-MINOS reconstruction, track matching, and selection 57.6\% of the time in the $\theta_\mu$ measurement range and 49.5\% in the $P_\mu$ range. These values receive contributions from muon acceptance between ArgoNeuT and MINOS, vertex reconstruction inefficiencies in ArgoNeuT, track reconstruction inefficiencies in both detectors, and selection efficiency. Inefficiencies due to acceptance arise from low-energy or large-angle muons that do not enter the active region of MINOS. A bin migration unfolding procedure is not applied to the reconstructed variables as no significant detector/reconstruction bias is present and the measurement resolution is finer than the bin width for all bins reported; the muon angular resolution over the majority of the measurement range is 1-1.5$^\circ$ and the momentum resolution is 5-10\% ~\cite{spitzthesis}. 


\begin{table}[t!]
\centering
\begin{tabular}{|c|c|c|}\hline 
E$_\nu$ bin (GeV) & \ Flux ($\nu$/GeV/m$^2$/10$^9$ POT)& \ Error \ \\  \hline\hline
0-1 & \ 8.3 $\times$ 10$^3$ & \  \ \\ 
1-2 & \ 4.3 $\times$ 10$^4$ & \ $^\dag$ \ \\ 
2-3 & \ 7.5 $\times$ 10$^4$ & \  \ \\ \hline
3-4 & \ 8.05 $\times$ 10$^4$ & \ 5.2 $\times$ 10$^3$ \ \\ \hline 
4-5 & \ 3.06 $\times$ 10$^4$ & \ 2.4 $\times$ 10$^3$ \ \\ \hline
5-7 & \ 9.07 $\times$ 10$^3$ & \ 5.3 $\times$ 10$^2$ \ \\ \hline
7-9 & \ 5.18 $\times$ 10$^3$ & \ 3.5 $\times$ 10$^2$ \ \\ \hline
9-12 & \ 3.21 $\times$ 10$^3$ & \ 2.2 $\times$ 10$^2$ \ \\ \hline
12-15 & \ 1.94 $\times$ 10$^3$ & \ 1.0 $\times$ 10$^2$ \ \\ \hline
15-18 & \ 1.09 $\times$ 10$^3$ & \ 65 \ \\ \hline
18-22 & \ 629 & \ 37 \ \\ \hline
22-26 & \ 348 & \ 20 \ \\ \hline
26-30 & \ 200 & \ 13 \ \\ \hline
30-36 & \ 119 & \ 6.8 \ \\ \hline
36-42 & \ 72.2 & \ 3.9 \ \\ \hline
42-50 & \ 51.6 & \ 2.8 \ \\ \hline
\end{tabular}
\centering
\caption{The neutrino flux corresponding to the differential cross section measurements. $^\dag$The fractional error on the 0-3~GeV range is conservatively set to 35\%.}
\label{fluxtable} 
\end{table}

\begin{figure}[h!]
\begin{center}
\begin{tabular}{c c}
\includegraphics[scale=.47]{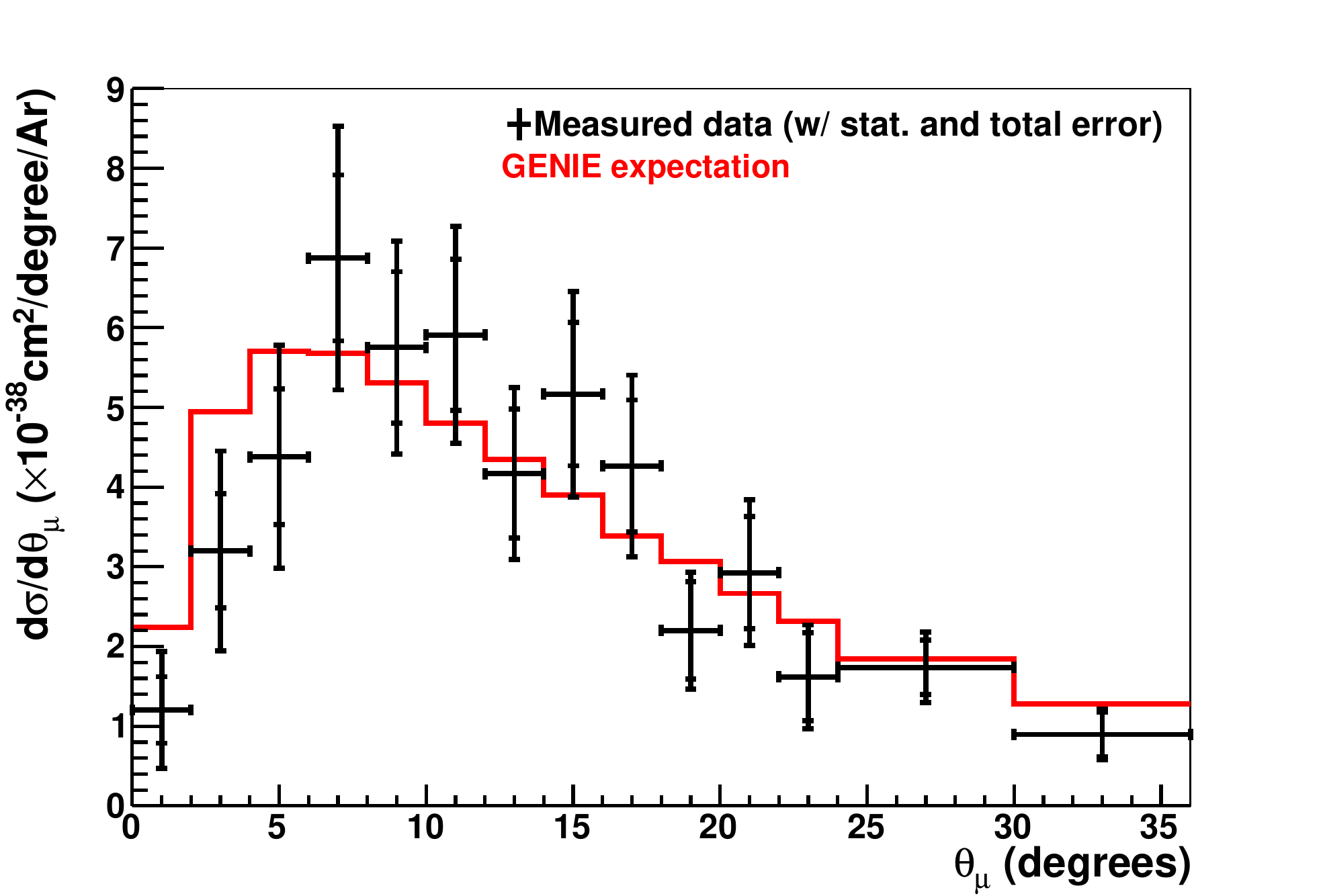}
\end{tabular}
\end{center}
\vspace{-.5cm}
\caption{The measured $\nu_{\mu}$ CC flux-integrated differential cross section (per argon nucleus) in muon angle.}
\label{fig:ardiff}
\end{figure}

\begin{figure}[h!]
\begin{center}
\begin{tabular}{c}
\includegraphics[scale=.47]{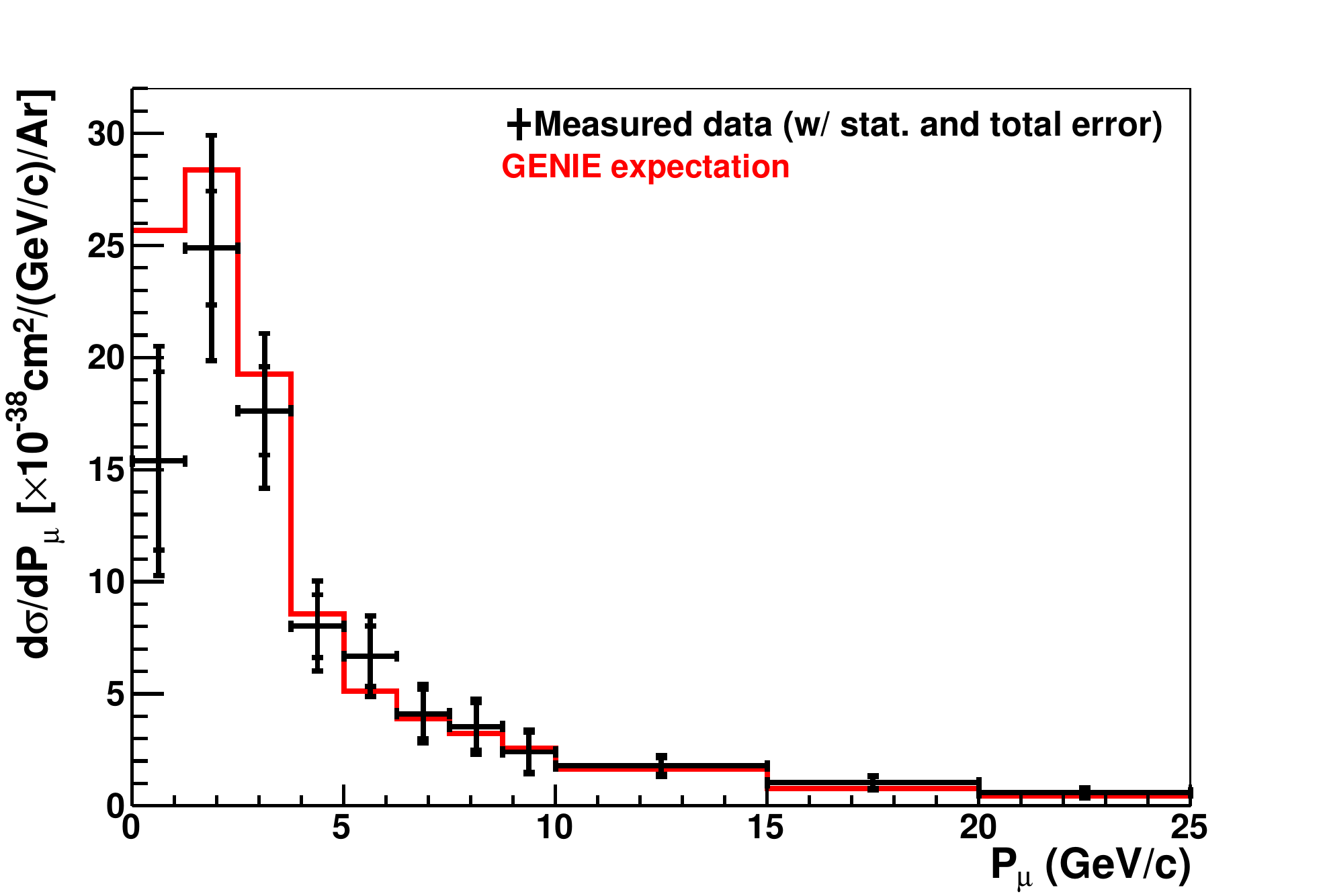}
\end{tabular}
\end{center}
\vspace{-.5cm}
\caption{The measured $\nu_{\mu}$ CC flux-integrated differential cross section (per argon nucleus) in muon momentum.}
\label{fig:ardiff2}
\end{figure}

The flux-integrated differential cross sections in $\theta_\mu$ and $P_\mu$ from $\nu_\mu$ CC events on an argon target are shown in Figs.~\ref{fig:ardiff} and \ref{fig:ardiff2}, respectively, and are tabulated in Tables~\ref{arcosmeas} and~\ref{armommeas}. The data and GENIE expectation agree well across most of the measurement ranges. More data are needed to confirm the apparent discrepancies at low angles and momenta. 

The differential cross section measurement uncertainties are dominated by statistics. The systematic error contributions are led by the 15.7\% uncertainty on the energy-integrated flux. Uncertainties associated with measurement resolution are evaluated by recalculating the differential cross sections after adjusting the measured $\theta_\mu$ and $P_\mu$ by $\pm$1$\sigma$, where $\sigma$ is the reconstructed variable's resolution. The uncertainty is conservatively set equal to the largest deviation from the central value, due to either the plus or minus 1$\sigma$ adjustment and the resulting bin weight redistribution. Other possible sources of systematic uncertainty have been found to be negligible.

\begin{table}[t!]
\centering
\begin{tabular}{|c|c|c|}\hline 
Measurement bin& \ $d\sigma/d\theta_{\mu}$ & Error\ \\
$[\theta_{\mu}]$ (degrees) & \ ($\frac{10^{-38}\mathrm{cm}^2}{\mathrm{degree}}$) & ($\frac{10^{-38}\mathrm{cm}^2}{\mathrm{degree}}$) \ \\  \hline\hline 
0-2 & \ 1.2 & 0.7\ \\ \hline 
2-4 & \ 3.2 & 1.3\ \\ \hline
4-6 & \ 4.4 & 1.4\ \\ \hline
6-8 & \ 6.9 & 1.7\ \\ \hline
8-10 & \ 5.8 & 1.3\ \\ \hline
10-12 & \ 5.9 & 1.4\ \\ \hline
12-14 & \ 4.2 & 1.1\ \\ \hline
14-16 & \ 5.2 & 1.3\ \\ \hline
16-18 & \ 4.3 & 1.1\ \\ \hline
18-20 & \ 2.2 & 0.7\ \\ \hline
20-22 & \ 2.9 & 0.9\ \\ \hline
22-24 & \ 1.6 & 0.7\ \\ \hline
24-30 & \ 1.7 & 0.4\ \\ \hline
30-36 & \ 0.9 & 0.3\ \\ \hline
\end{tabular} 
\caption{The measured $\nu_{\mu}$ CC flux-integrated differential cross section (per argon nucleus) in muon angle.}
\label{arcosmeas}
\end{table}
\begin{table}[t!]
\centering
\begin{tabular}{|c|c|c|}\hline 
Measurement bin& \ $d\sigma/dP_{\mu}$ & Error\ \\
$[P_{\mu}]$ (GeV/c) & \ ($\frac{10^{-38}\mathrm{cm}^2}{\mathrm{GeV/c}}$) & ($\frac{10^{-38}\mathrm{cm}^2}{\mathrm{GeV/c}}$) \ \\  \hline\hline 
0.00-1.25 & \ 15.4 & 5.1\ \\ \hline 
1.25-2.50 & \ 24.9 & 5.0\ \\ \hline
2.50-3.75 & \ 17.6 & 3.5\ \\ \hline
3.75-5.00 & \ 8.0 & 2.0\ \\ \hline
5.00-6.25 & \ 6.7 & 1.8\ \\ \hline
6.25-7.50 & \ 4.1 & 1.3\ \\ \hline
7.50-8.75 & \ 3.5 & 1.2\ \\ \hline
8.75-10.0 & \ 2.4 & 1.0\ \\ \hline
10.0-15.0 & \ 1.8 & 0.5\ \\ \hline
15.0-20.0 & \ 1.0 & 0.3\ \\ \hline
20.0-25.0 & \ 0.6 & 0.2\ \\ \hline
\end{tabular} 
\caption{The measured $\nu_{\mu}$ CC flux-integrated differential cross section (per argon nucleus) in muon momentum.}
\label{armommeas}
\end{table}


Differential cross sections on an isoscalar target are useful for a simple comparison of these results to other measurements on different nuclei. The correction for transforming the argon target measurement reported here into an isoscalar one is arrived at by reweighting each GENIE simulated $\nu_\mu$ CC interaction based on its nucleon target. The extracted multiplication factor of 0.96 can be applied to each on-argon differential cross section measurement bin in order to obtain the differential cross sections on an isoscalar target. This correction factor is model-dependent as it relies on GENIE's underlying neutrino cross section predictions for the various interaction channels.

Although comparing the differential cross section results to other measurements is difficult, the total cross section can be extracted with the complete sample size (379~events), background expectation (18~events), and overall detection efficiency (49.5\%). The total cross section includes the bins outside of the $\theta_\mu$ and $P_\mu$ differential cross section measurement ranges. The measured total $\nu_\mu$ CC cross section is $\sigma/E_\nu$=(7.3$\pm$1.2)$\times10^{-39}\frac{\mathrm{cm}^2}{\mathrm{GeV}}$ per isoscalar nucleon at $\langle E_\nu \rangle=4.3$~GeV, consistent with the most precise total cross section measurements available at these energies~\cite{flux,nomadcc}. The argon-to-isoscalar correction has been applied in arriving at this value.

The first two weeks of the ArgoNeuT physics run, comprising the entirety of the neutrino-mode data acquisition, have been analyzed. An additional 1.25$\times10^{20}$~POT taken in anti-neutrino mode over the 5.5~month physics run is currently being analyzed. LArSoft reconstruction, ArgoNeuT-MINOS track matching efficiency, and measurement resolution will be augmented for future analyses. Subsequent work will demonstrate the complete power of LArTPC technology with $\frac{dE}{dx}$-based particle identification and calorimetry in general. 

ArgoNeuT has performed the first $\nu_\mu$ CC differential cross section measurements for scattering on argon. The results are consistent with the GENIE neutrino event generator predictions from $0^\circ$$<\theta_\mu$$<36^\circ$ and $0$$<P_\mu$$<25$~GeV/c and the measured total $\nu_\mu$ CC cross section at $\langle E_\nu \rangle=4.3$~GeV is consistent with the world's data. The differential cross sections elucidate the behavior of the outgoing muon in $\nu_\mu$ CC interactions, information useful for tuning neutrino event generators, reducing the systematics associated with a long baseline neutrino oscillation experiment's near-far comparison, and informing the theory of the neutrino-nucleus interaction in general. In addition to importance in understanding neutrino scattering and relevance for neutrino oscillations, these measurements represent a significant step forward for LArTPC technology as they are among the first with such a device.

We gratefully acknowledge the cooperation of the \newline MINOS collaboration in providing their data for use in this analysis. We wish to acknowledge the support of Fermilab, the Department of Energy, and the National Science Foundation in ArgoNeuT's construction, operation, and data analysis.
\bibliographystyle{apsrev}   
\bibliography{ccincbib}
\end{document}